# Stoichiometry-Driven Switching between Surface Reconstructions on SrTiO$_3$(001)

Stefan Gerhold[*], Zhiming Wang, Michael Schmid, and Ulrike Diebold

Institute of Applied Physics, Vienna University of Technology

## Abstract

Controlling the surface structure on the atomic scale is a major difficulty for most transition metal oxides; this is especially true for the ternary perovskites. The influence of surface stoichiometry on the atomic structure of the SrTiO$_3$(001) surface was examined with scanning tunneling microscopy, low-energy electron diffraction, low-energy He$^+$ ion scattering (LEIS), and X-ray photoelectron spectroscopy (XPS). Vapor deposition of 0.8 monolayer (ML) strontium and 0.3 ML titanium, with subsequent annealing to 850 °C in 4×10$^{-6}$ mbar O$_2$, reversibly switches the surface between c(4×2) and (2×2) reconstructions, respectively. The combination of LEIS and XPS shows a different stoichiometry that is confined to the top layer. Geometric models for these reconstructions need to take into account these different surface compositions.

[*] Corresponding author
 *Tel.*: +43-(0)1-58801-13483.
 *E-mail address*: gerhold@iap.tuwien.ac.at
 *Full postal address*: Institute of Applied Physics (E134), Vienna University of Technology, Wiedner-Hauptstraße 8-10, 1040 Wien, Austria.

# 1. Introduction

Strontium titanate (SrTiO$_3$, STO) belongs to the class of cubic perovskite oxides and it is well known for its useful bulk and surface properties, e.g., a high dielectric constant at low temperatures [1], photocatalytic water-splitting [2], lattice matching for growth of high-T$_c$ superconductors [3], and the formation of two-dimensional electron gases at its surfaces [4,5], and at interfaces with other perovskites [6]. For most applications, a control of the surface structure at the atomic scale is of central importance.

The most important surface, STO(001), exhibits a large variety of reconstructions. Typical preparation procedures consist of sputtering with Ar$^+$ ions of different energies (~1 keV) and fluences, and annealing to high temperatures (~1000 °C) in various oxygen pressures (atmospheric to <10$^{-10}$ mbar O$_2$). Often the outcome also depends on the sample preparation history. Different groups report numerous STO(001) surface reconstructions, including (1×1) [7-13], (2×1) [9-12,14,15], (2×2) [7,9,10,16], c(4×2) [12,14,16,17], c(4×4) [12,16], (4×4) [16], c(6×2) [11,15,17-19], (√5×√5)-R26.6° [13,16,20,21], and (√13×√13)-R33.7° [11,16]. Sometimes different reconstructions are reported for very similar preparation conditions.

At the STO(110) surface, a study by Wang et al. [22] showed that a series of reconstructions can be created and converted into each other *reversibly* by controlling the surface stoichiometry, i.e., by depositing Sr or Ti onto the surface and annealing in 10$^{-6}$ mbar O$_2$. This was recently also demonstrated at the STO(111) surface by Feng et al. [23]. Following this approach, the present Letter reports on the response of STO(001) surface reconstructions to a similar procedure. Using low-energy electron diffraction (LEED), scanning tunneling microscopy (STM), X-ray photoelectron spectroscopy (XPS), and low-energy He$^+$ ion scattering (LEIS) it is concluded that the transition between the (2×2) and the c(4×2) STO(001) surface reconstructions is driven by surface stoichiometry. The implications for models proposed in the literature are discussed.

## 2. Experimental Methods

Nb-doped (0.5 wt %) SrTiO$_3$(001) single crystals were purchased from MaTecK Company, Germany. After ultrasonic cleaning in acetone, the samples were introduced into a two-chamber UHV system. One chamber is equipped with evaporators (Sr, Ti), a sputter gun, a home-built quartz crystal microbalance (QCM) and a leak valve for admitting O$_2$ into the chamber. Its base pressure was below 10$^{-9}$ mbar. The second chamber, with a base pressure below 10$^{-10}$ mbar, was used for analysis with LEED (SpectaLEED, Omicron), STM (Aarhus 150, SPECS), XPS (non-monochromatized dual-anode), and a scanning ion gun for LEIS. XPS spectra were acquired using Mg K$\alpha$ radiation. For LEIS, 1 keV He$^+$ ions were used. Backscattered ions (scattering angle $\vartheta=137°$) and photoelectrons (emission normal to sample surface) were detected with a SPECS Phoibos 100 hemispherical analyzer with 5-channel detection. The XPS peaks were fitted after subtracting a Shirley background and the ISS peak intensity was summed up over the peak area after subtraction of a linear background.

After loading the samples into the system, they were sputtered with 1 keV Ar$^+$ ions (~8×10$^{13}$ Ar$^+$/cm$^2$s) for typically 10 minutes and annealed at ~850 °C in 10$^{-6}$ mbar O$_2$ for 40 minutes. Titanium was evaporated from an electron beam evaporator (Omicron) and strontium was evaporated using a low-temperature effusion cell (CreaTec). The flux was monitored by a QCM. The temperature was measured with an optical pyrometer using an emissivity of 1. While this measurement method is not very accurate for oxides, the structures presented here are stable within a wide range of temperatures (700-900 °C), thus ambiguities in temperature measurement should not be a major a problem.

All samples were treated with the following sequence. In order to get a clean, well-defined surface structure, samples introduced to the UHV system were sputtered (at least once) and then annealed. A c(4×2) structure (determined by LEED or STM) was observed after this procedure. Sr was evaporated onto this surface and the sample was annealed after deposition. Sr was deposited until a change of surface reconstruction to the (2×2) structure was observed (by LEED). For reversing the reconstruction, Ti was deposited until the structure changed back to the original surface reconstruction [c(4×2)]. A full switch of reconstructions was obtained by depositing a nominal thickness (determined by the QCM) of ~3 Å of Sr or ~0.35 Å of Ti. Taking the densities of Ti and Sr and the size of the STO(001) unit cell into account, this corresponds to 0.3 and 0.8 monolayers (MLs) for Ti and Sr, respectively. After deposition, the samples were annealed for 15 minutes (850 °C in $10^{-6}$ mbar $O_2$).

## 3. Results

A two-domain c(4×2) reconstructed surface was obtained through multiple sputtering and annealing cycles; no other superstructures were detected by STM and LEED. Figure 1a) shows a large-scale STM image of this surface. The inset shows the associated LEED pattern. This structure was identified as the two-domain c(4×2) reconstruction using the software LEEDpat30 [24]. The terraces are ~30 nm wide and the step-height is equivalent to one unit cell of STO (3.9 Å). An atomically resolved image of the two-domain c(4×2) structure and its corresponding fast Fourier transformation is shown in Figure 1(a) with unit cells indicated.

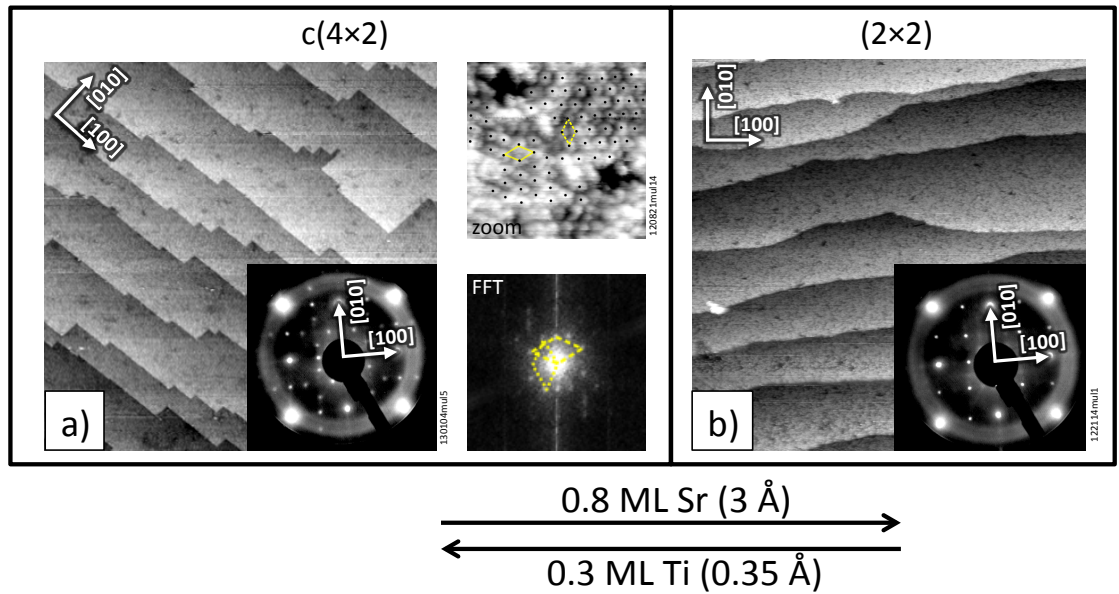

Figure 1: Large-scale STM images (200×200 nm$^2$) of differently reconstructed STO(001) surfaces. The insets show the LEED patterns at an electron energy of 69 eV. (a) Multiple sputtering and annealing produces the two-domain c(4×2) surface (tunneling parameters: $V_s$=+3.5 V, $I_t$=0.48 nA). An atomically resolved STM image and the corresponding FFT are shown at the side of (a). (b) Deposition of Sr with subsequent annealing changes the reconstruction to a (2×2) structure (tunneling parameters: $V_s$=+2.3 V, $I_t$=0.26 nA). Ti deposition and annealing reverses the surface back to the c(4×2) structure. Note that the step edges are different in (a) and (b).

One prominent feature of this surface are the straight step edges that run along the [010] and [100] directions. On the c(4×2) reconstructed surface, step edges aligned with the <100> directions have also been reported by Jiang and Zegenhagen [17] and Castell [12]. Deposition of 0.8 ML (3 Å QCM reading) of Sr onto this surface and annealing to 850 °C in 10$^{-6}$ mbar O$_2$ changes the surface (Figure 1 b). The LEED pattern reveals a (2×2) structure (inset). Again the terraces are ~30 nm wide with one unit cell step-height. The sharp LEED pattern indicates that the structure covers the surface uniformly and with good order for the most part. However, faint intermediate spots in the LEED image indicate the presence of a minority phase with different periodicity. Note the different appearance of the step edges; for the (2×2) structure the step edges are wavy, as was also found by Silly et al. [25]. Because it was not possible to image the (2×2) structure atomically, UHV flash-annealing (~850 °C, 5 min) was applied (to increase the conductivity).

However, the surface transformed into a structure with c(4×4) symmetry (not shown). This transition was already reported by Silly et al. [25] and the structural appearance in STM exhibited also the same c(4×4) structure, as was found in their work.

The process of transforming the surface structure can be reversed; switching between c(4×2) and (2×2) reconstructions was reproduced more than ten times. As indicated by the arrow in the bottom of Figure 1, deposition of Ti onto the (2×2) structure and annealing at ~850 °C in $10^{-6}$ mbar $O_2$ leads back to the two-domain c(4×2) structure. The amount of Ti necessary to switch between the two structures was measured to be 0.3 ML (~0.35 Å QCM reading). This procedure yields sharp LEED patterns, no degradation of the surface quality (as judged by LEED) was observed for multiple switchings.

To determine the chemical composition of the surface, XPS and LEIS were performed. Figure,2 shows the XPS and LEIS spectra for the two-domain c(4×2) and the (2×2) reconstructed surface, respectively. Well-defined O1$s$, Ti2$p$, and Sr3$d$ core-level spectra were observed. The Ti2$p$ spectrum shows the single doublet feature of the Ti$^{4+}$ state as in bulk STO. Furthermore, there is no detectable feature related to hydroxyls in O1$s$ spectra, indicating that those samples were fully oxidized and not affected by, e.g., adsorbed water. All peaks of the c(4×2) reconstructed surface have a binding energy that is 0.11 eV higher than those of the (2×2) reconstructed surface. This indicates downward band bending on the c(4×2) surface, i.e., a surface that is more positively charged. The ratio of the intensities of the Sr3$d_{5/2}$ to Ti2$p_{3/2}$ peaks was determined in order to reveal compositional differences. Within the statistical error, the XPS signal (in normal emission) did not show any difference (see table 1). Note that the escape depth of photoelectrons in normal emission is rather large (1.2 to 1.7 nm as calculated with the SESSA code [26]), thus XPS is not very sensitive to the composition at the very surface).

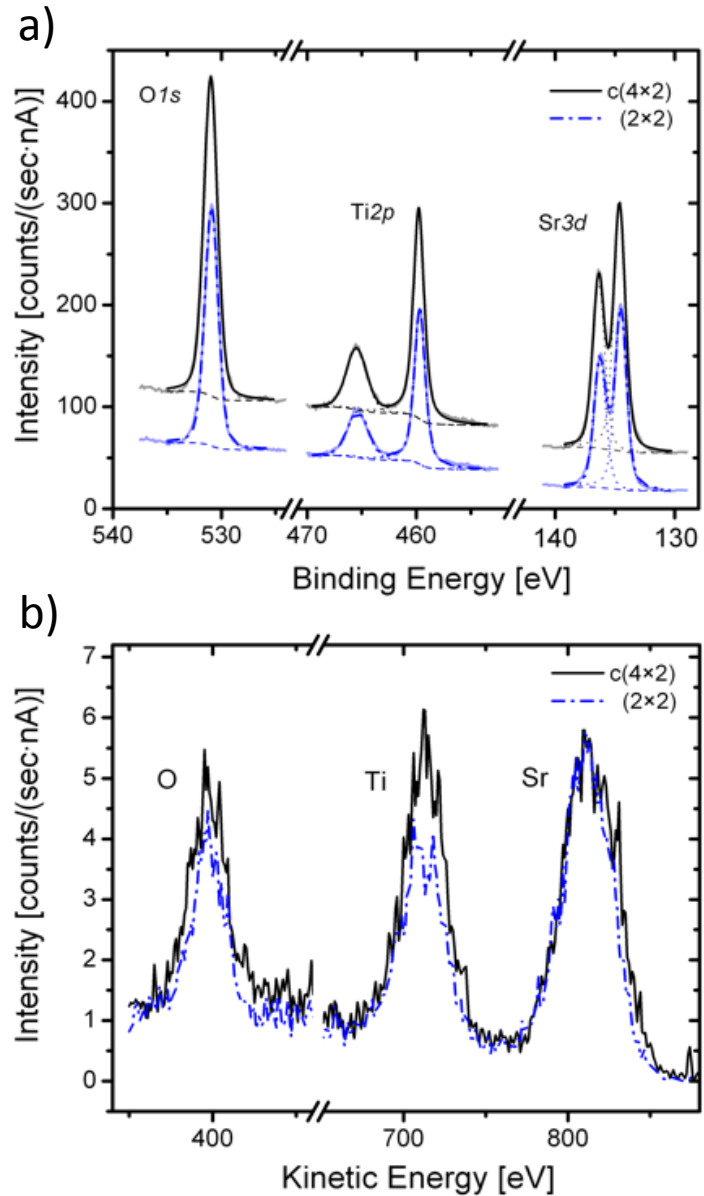

Figure 2: (a) XPS spectra (Mg K$_\alpha$ with hv=1254 eV, normal emission) and (b) LEIS spectra (averaged over 10 scans) (He$^+$ with 1 keV, scattering angle θ=137°) of SrTiO$_3$(001) (2×2) and c(4×2) reconstructed surfaces. The Sr to Ti peak intensity ratios are listed in Table 1.

Figure 2(b) shows the spectra for both surface reconstructions obtained with LEIS, which is a very surface sensitive technique. The spectra are averaged over 10 scans. The ratios of the intensities of Sr to Ti peaks are listed in Table 1. Clearly, a higher strontium concentration is visible for the (2×2) reconstructed surface.

| Intensity Sr/Ti | (2×2) | c(4×2) |
|---|---|---|
| XPS | 1.03 ± 0.02 | 1.06 ± 0.02 |
| LEIS | 1.70 ± 0.20 | 1.29 ± 0.15 |

Table1: Ratio of Sr and Ti peak intensities of the two different reconstructions measured with LEIS and XPS. The ratio was calculated by integrating the intensities of the characteristic peaks in XPS ($Sr3d_{5/2}$ to $Ti2p_{3/2}$; after subtracting a Shirley background) and LEIS (Sr to Ti; after subtracting a linear background) spectra.

## 4. Discussion

Reconstructions of STO(001) surfaces have been proposed to be formed by ordered oxygen vacancies (on vacuum-annealed samples) [11,27], Sr adatoms [16,19] or a double-layer $TiO_2$ structure [28-30]. For the reconstructions investigated here, mainly two structure proposals are found in the literature. Based on STM measurements and first-principles calculations, Kubo and Nozoye suggested a model consisting of ordered Sr adatoms [16]. Supported by a combined STM and density functional theory (DFT) study, Marks and coworkers suggested a double-layer $TiO_2$ structure forming both, the c(4×2) [29] and the (2×2) structure [30]. The two structures are inherently similar, i.e., shifting every second row of the (2×2) structure by one unit cell of STO results in the creation of the c(4×2) structure. Therefore the net stoichiometry per surface unit cell for both model types is equal for the c(4×2) and the (2×2) reconstruction. The transformation between these structures upon Sr and Ti deposition as well as our LEIS result indicate, however, that the two structures show a different stoichiometry that is confined to the surface layer. For comparison, we performed ion scattering on the well-known, tetrahedrally coordinated $TiO_2$-terminated (4×1) and (5×1) reconstructed STO(110) surface (consisting of a single TiO$_2$-like layer) (not shown here). The spectra showed a significantly lower Sr/Ti peak intensity ratio, increasing to values similar to the present ones only with increasing ion beam damage. Thus, the LEIS results are incompatible with purely TiO$_2$-terminated surfaces for SrTiO$_3$(001)-(2×2) and c(4×2). With STM we saw no sign of phase separation [31] that could give rise to the Sr signal in LEIS.

Becerra-Toledo et al. discussed the role of water (in the residual gas during annealing) on the stability of STO(001) surfaces [32]. They computed surface energies of STO(001) surface reconstructions and found that dissociatively adsorbed water is able to stabilize different surface structures [(2×1), c(4×4), and c(4×2)]. As an indication of hydroxyl species on the surface, the low binding-energy (BE) shoulder of the Ti2$p$ 4+ peaks and the high-BE shoulder of the O1$s$ peak was analyzed in their work. Figure 2(a), however, does not show any sign of Ti$^{3+}$ or OH, indicating that the present structures are fully oxidized and stable without dissociatively adsorbed water.

A striking feature of the two surfaces is the different appearance of the step edges. As already discussed in references [12,17,25] the c(4×2) surface usually shows straight step edges aligned with the <001> directions while the (2×2) surface shows wavy step edges. Fompeyrine et al. investigated the STO(001) surface with friction force microscopy and reported that SrO terminated layers show wavy step edges while TiO$_2$ terminated layers exhibit straight and aligned step edges [33]. In any case, the different appearance of steps indicates that the building blocks of the two surfaces should be different.

The step heights in the large-scale STM images (Figure 1) as well as the similar appearance of the structure everywhere on the surface show that the present surfaces had a single type of termination. In the simplest model conceivable, the evaporation of Sr formed half a monolayer of STO(001) and switched the termination from TiO$_2$ to SrO. Such a model is not realistic, however, as the reconstructions are certainly more complex than a simple modification of the $TiO_2$ or SrO bulk termination; furthermore, the amount of 0.3 ML Ti is not enough for this change. The amount of Sr needed for the conversion in the reverse direction is larger by a factor of 2.7, which is astonishing when considering the 1:1 ratio of Sr and Ti in the bulk (repeated switching back and forth by successive Ti and Sr deposition is possible only if the excess material eventually forms additional $SrTiO_3$ bulk layers). At this point it can only be speculated whether the sticking coefficient of Sr and Ti is different by this amount, whether some Sr desorbs upon annealing, or Ti diffusion from the bulk plays a role.

Based on the study of Wang et al. [22] and the results presented here, the surface stoichiometry is considered the underlying reason behind the creation of different reconstructions. From the viewpoint of thermodynamic stability, the surface free energy of a reconstructed surface can be expressed in terms of the chemical potentials of the involved species (compare the study of Li et al. [34]). By varying the chemical potential of Ti or Sr (deposition of Ti or Sr), the surface will undergo a transition to a different geometric structure and therefore lower its free energy.

## **Conclusion**

In conclusion, the present study shows that, on the STO(001) surface, structures can be changed reversibly by depositing Sr or Ti onto the surface and annealing in $O_2$ environment. XPS and LEIS measurements indicate that the (2×2) and c(4×2) structures show a different stoichiometry, which is confined to the surface layer. Evaporation of a material increases its chemical potential and it therefore alters the free energy of the surface. The surface stoichiometry is considered the underlying reason behind the creation of different reconstructions. Valid structural models have to take the different stoichiometry of these structures into account, as well as the fact that the LEIS results are incompatible with a pure $TiO_2$ termination. However, the models for the (2×2) and c(4×2) structures proposed in the literature so far are inherently similar. These models incorporate the same building block with an equal density for the two structures. This is in contrast to the results presented here. Assuming the structure for the c(4×2) reconstruction proposed in ref. [29] is correct, at least a new model for (2×2) has to be found.

## **Acknowledgement**

This work was supported by the Austrian Science Fund (FWF, project F45).